\documentclass[prb,amsmath,twocolumn,showpacs,superscriptaddress]{revtex4-1}

\usepackage{bm}
\usepackage{graphicx}
\usepackage[usenames,dvipsnames,svgnames,table]{xcolor}
\usepackage[colorlinks=true,linkcolor=RoyalBlue,citecolor=RoyalBlue]{hyperref}
\usepackage{cleveref}

\newcommand{\ket}[1]{|#1\rangle}
\newcommand{\bracket}[1]{\langle #1 \rangle}

\newcommand{\eps}{\varepsilon}

\begin{document}

\title{Strain-Fluctuation-Induced Near-Quantization of Valley Hall Conductivity in Graphene Systems}
\author{Wen-Yu Shan}
\affiliation{Department of Physics, School of Physics and Electronic Engineering, Guangzhou University, Guangzhou 510006, China}
\affiliation{Department of Physics, Carnegie Mellon University,
  Pittsburgh, Pennsylvania 15213, USA}

\author{Di Xiao}
\affiliation{Department of Physics, Carnegie Mellon University,
  Pittsburgh, Pennsylvania 15213, USA}

\date{\today}

\begin{abstract}
  We develop a theory of the valley Hall effect in high-quality
  graphene samples, in which strain fluctuation-induced random gauge
  potentials have been suggested as the dominant source of disorder.
  We find a near-quantized value of valley Hall conductivity
  in the band transport regime, which originates from an enhanced side
  jump of a Dirac electron when it scatters off the gauge potential.
  By assuming a small residue charge density our theory reproduces qualitatively
  the temperature- and gap-dependence of the observed valley Hall
  effect at the charge neutral point.  Our study suggests that the
  valley Hall effect in graphene systems represents a new paradigm for
  the anomalous Hall physics where gauge disorder plays an important
  role.
 
\end{abstract}

\maketitle

\section{Introduction}

Charge carriers in graphene can be described by the two-dimensional (2D) Dirac equation, which exhibit a slew of interesting electronic properties~\cite{castro2009}.  One of the consequences is that strains behave as pseudo-magnetic fields for Dirac electrons, and carefully designed lattice deformation pattern can result in the formation of Landau levels~\cite{guinea2010}.  Even without the engineered strains, random strain fluctuations are inevitable in 2D materials~\cite{meyer2007,fasolino2007,vozmediano2010,amorim2016}.  They appear in the form of either out-of-plane corrugations due to thermal ripples or in-plane displacement from the interaction with substrates.  Strain fluctuations then act like random magnetic (gauge) field, and can significantly affect transport behaviors.  Their effect has been extensively explored in the longitudinal transport phenomena~\cite{ludwig1994,couto2014,engels2014,ochoa2013,vicent2017}, such as weak localization~\cite{couto2014,engels2014} and spin relaxation phenomena~\cite{ochoa2013,vicent2017}.

In this work, we investigate the role of random strain fluctuations on a particular type of transverse transport phenomena---the valley Hall effect of 2D Dirac electrons~\cite{xiao2007}.  The valley Hall effect can be regarded as two opposite copies of the anomalous Hall effect of a pair of gapped Dirac points related by time-reversal symmetry~\cite{nagaosa2010}.  That is, carriers in the two valleys will flow in the opposite transverse direction upon the application of a longitudinal electric field.  Our motivation is twofold. First, in gapped monolayer and bilayer graphene systems, the valley Hall effect has been observed experimentally~\cite{gorbachev2014,sui2015,shimazaki2015}.  Thus, detailed experimental study of Hall-type transport in the presence of strain fluctuations is feasible.  Secondly, all recent valley Hall measurements in graphene are carried out in high-quality devices in which strain fluctuations are the dominant source of disorder~\cite{couto2014,engels2014}.  However, no existing theories have discussed its effect on the Hall transport. We will show that, the strain fluctuations are essential to understand the valley Hall effect of 2D Dirac electrons, and provide a new insight to recent debates~\cite{lensky2015,kirczenow2015,li2011,zhu2017,song2018,brown2018} on the observed nonlocal signals at the charge neutral point~\cite{gorbachev2014,sui2015,shimazaki2015}.

Our main results are summarized below.  Focusing on the band transport
regime, we find that the valley Hall conductivity exhibits a singular
behavior in the presence of strain fluctuation-induced long-range gauge disorder: as the Fermi
level sweeps across the band edge, it jumps from zero to a
\emph{nearly} quantized value, $2e^2/h$ for monolayer graphene and
$4e^2/h$ for bilayer graphene.  The origin of this singular behavior
is traced back to an enhanced side jump of a Dirac electron when it
scatters off the gauge potential~\cite{yang2011,yang2011a}.
Furthermore, at the charge neutral point, by assuming a small residue
charge density we calculate the temperature- and gap-dependence of the
valley Hall conductivity, which qualitatively agrees with the
experiment~\cite{sui2015}.  We also find that strain-induced
long-range scalar potential can reduce the valley Hall conductivity
from its quantized value.  Our study suggests that the valley Hall
effect in graphene systems represents a new paradigm for the anomalous
Hall physics where gauge disorder plays an important role.  Our theory can also be applied to other 2D valley Hall materials such as transition metal 
dichalcogenides as well~\cite{wu2018,hung2019}.

The paper is organized as follows. In Sec.~\ref{sec:intrinsic} we present the intrinsic valley Hall effect. In Sec.~\ref{side_jump} we study the random strain-induced side jump and the resulting valley Hall effect. Temperature dependence and rigorous numerical analysis are shown in Sec.~\ref{temperature} and Sec.~\ref{numerics}, respectively. In Sec.~\ref{scalar}, effect of long-range scalar potential is investigated. Finally, discussion and conclusion are made in Sec.~\ref{conclusion}. Technical details are relegated to the appendixes.

\section{Intrinsic valley Hall effect}\label{sec:intrinsic}

We begin with the following effective Hamiltonian
\begin{equation} \label{ham} 
H_0 = \hbar v\bm k \cdot \bm\sigma + \Delta\sigma_z \;,
\end{equation}
which describes the low-energy electron dynamics in one of the Dirac
valleys in gapped graphene.  Here $v$ is the velocity,
$\bm k = (k_x, k_y)$ is the two-dimensional wave vector, $2\Delta$ is
the band gap opened by inversion symmetry breaking, and $\bm\sigma$
represents the sublattice indices.  The Hamiltonian for the other
valley can be obtained by performing a time-reversal operation on
$H_0$.  The energy dispersion is given by
$\eps_{c,v} = \pm \eps_{\bm k} = \pm (\hbar^2v^2k^2 + \Delta^2)^{1/2}$
with the corresponding eigenstates
\begin{equation} \label{effH}
\ket{u_{\bm k}^c} = \binom{\cos\frac{\theta_{\bm k}}{2}}{\sin\frac{\theta_{\bm k}}{2}e^{i\phi_{\bm k}}} \;, \quad
\ket{u_{\bm k}^v} = \binom{\sin\frac{\theta_{\bm k}}{2}e^{-i\phi_{\bm k}}}{-\cos\frac{\theta_{\bm k}}{2}} \;,
\end{equation}
where the superscript $c$ and $v$ label the conduction and valence
bands, respectively, and the angular variables $\theta_{\bm k}$ and
$\phi_{\bm k}$ are defined as
$\theta_{\bm k} \equiv \cos^{-1}(\Delta/\eps_{\bm k})$ and
$\phi_{\bm k} \equiv \tan^{-1}(k_y/k_x)$.  The opening of the band gap
gives rise to nonzero Berry curvature~\cite{xiao2007}, defined by
$\bm\Omega_n(\bm k) = \Omega_n(\bm k) \hat z = i\bracket{\nabla_{\bm
    k}u_{\bm k}^n|\times|\nabla_{\bm k}u_{\bm k}^n}$~\cite{xiao2010}.
For the two-band model given in Eq.~\eqref{ham} we have
\begin{equation}
\Omega_c(\bm k) = -\Omega_v(\bm k) = -\frac{\hbar^2v^2\Delta}{2(\hbar^2v^2k^2+\Delta^2)^{3/2}} \;.
\end{equation}
The Berry curvature in the other valley has opposite sign, as required
by time-reversal symmetry.

Assuming weak inter-valley scattering, we can decouple the valley Hall
effect into two copies of the anomalous Hall effect for each valley
species.  However, this decoupling must be treated with care.  In the
anomalous Hall effect, there is an intrinsic contribution to the Hall
conductivity, given by the summation of the Berry curvature over all
occupied Bloch states~\cite{nagaosa2010}.  It can be divided into two
parts.  One comes from fully occupied bands.  This part manifests as
chiral edge states at the Fermi energy, and gives rise to a quantized
contribution to the Hall conductivity.  On the other hand, the valley
Hall systems considered here are topologically trivial without
protected edge states.  Therefore no electronic transport is possible
when the Fermi energy is inside the band gap~\footnote{Specifically,
  if one integrates the Berry curvature over the valence band of the
  gapped Dirac Hamiltonian, one would obtain a Hall conductivity of
  $\pm e^2/h$ for the two valleys (counting the spin degeneracy).
  Naively taking their difference seems to suggest that the valley
  Hall conductivity should be $2e^2/h$ when the Fermi energy is in the
  band gap.  However, since there is no topological edge states, this
  result is inconsistent with the general notion that electronic
  transport should be a Fermi surface property and fully occupied
  topologically trivial bands should not contribute to electronic
  transport.  The nonzero result is an artifact due to treating the
  two Dirac valleys separately, which cannot capture the global
  topology of the bands.}.  Consequently, we shall drop this
contribution in the calculation of the valley Hall conductivity.  This
leaves us with the contribution from partially occupied bands.
Haldane has argued that this contribution can be written as the Berry
phase of quasiparticles moving on the Fermi surface, and thus it can
be regarded as a Fermi surface property~\cite{haldane2004}.  For an
electron-doped sample, the Fermi surface contribution is
\begin{equation}\label{intrins}
\sigma_H^\text{int} = 4\frac{e^2}{\hbar}\sum_{\bm k} \Omega_c(\bm k) \Theta(\eps_F - \eps_{\bm k}) = \frac{2e^2}{h}(1 - \cos\theta_F) \;,
\end{equation}
where the factor of 4 counts the spin and valley degeneracy, $\eps_F$
is the Fermi energy, and $\theta_F \equiv \theta_{k_F}$.  When
$\eps_F$ approaches the band edge, $\sigma_H^\text{int}$ vanishes.  If
the sample is hole doped, then one should consider the Fermi surface
contribution of holes, which is opposite to that of electrons.
Obviously, the intrinsic contribution alone cannot explain the
observed valley Hall effect around the charge neutral
point~\cite{gorbachev2014,sui2015,shimazaki2015}.

\section{Random strain-induced side jump}\label{side_jump}

To remedy this situation, we consider the effect of random
strain-induced gauge disorder.  In graphene systems, strain can be
induced either by out-of-plane corrugations or by in-plane
displacements of the carbon atoms.  Both will generate a random gauge
potential for Dirac electrons~\cite{vozmediano2010},
\begin{equation} \label{gauge_dis}
V_\text{imp}(\bm r) = d_x(\bm r) \sigma_x + d_y(\bm r) \sigma_y \;.
\end{equation}
The point group symmetry of graphene requires that $\bm d(\bm
r)\propto(u_{xx}-u_{yy},-2u_{xy})$, where $u_{\alpha\beta}$ is the
strain tensor defined in terms of the deformation field $\bm u(\bm r)$,
$u_{\alpha\beta} = (\partial_\alpha u_\beta + \partial_\beta u_\alpha)/2$.
The Fourier transform of $\bm d(\bm r)$ has the form~\cite{couto2014}
\begin{equation} \label{dq}
d_{\pm}(\bm q)= F(\bm q) q^2 e^{\mp 2i\varphi_{\bm q}} \;,
\end{equation}
where $d_{\pm}(\bm q)=d_x(\bm q)\pm i d_y(\bm q)$, and $F(\bm q)$ is a
prefactor depending on the details of the strain field.  The
appearance of the phase angle $2\varphi_{\bm q}$, defined by
$\varphi_{\bm q}\equiv\tan^{-1}(q_y/q_x)$, is due to the fact that
$u_{\alpha\beta}$ is a second-order derivative of either the height
field (out-of-plane corrugations) or random potentials from substrates
(in-plane displacements)~\cite{couto2014,vozmediano2010}.  Both modes
are long-wavelength elastic modes, as indicated by the
$\bm q$-dependence of $F(\bm q)$.

\begin{figure}[t]
\centering \includegraphics[width=0.43\textwidth]{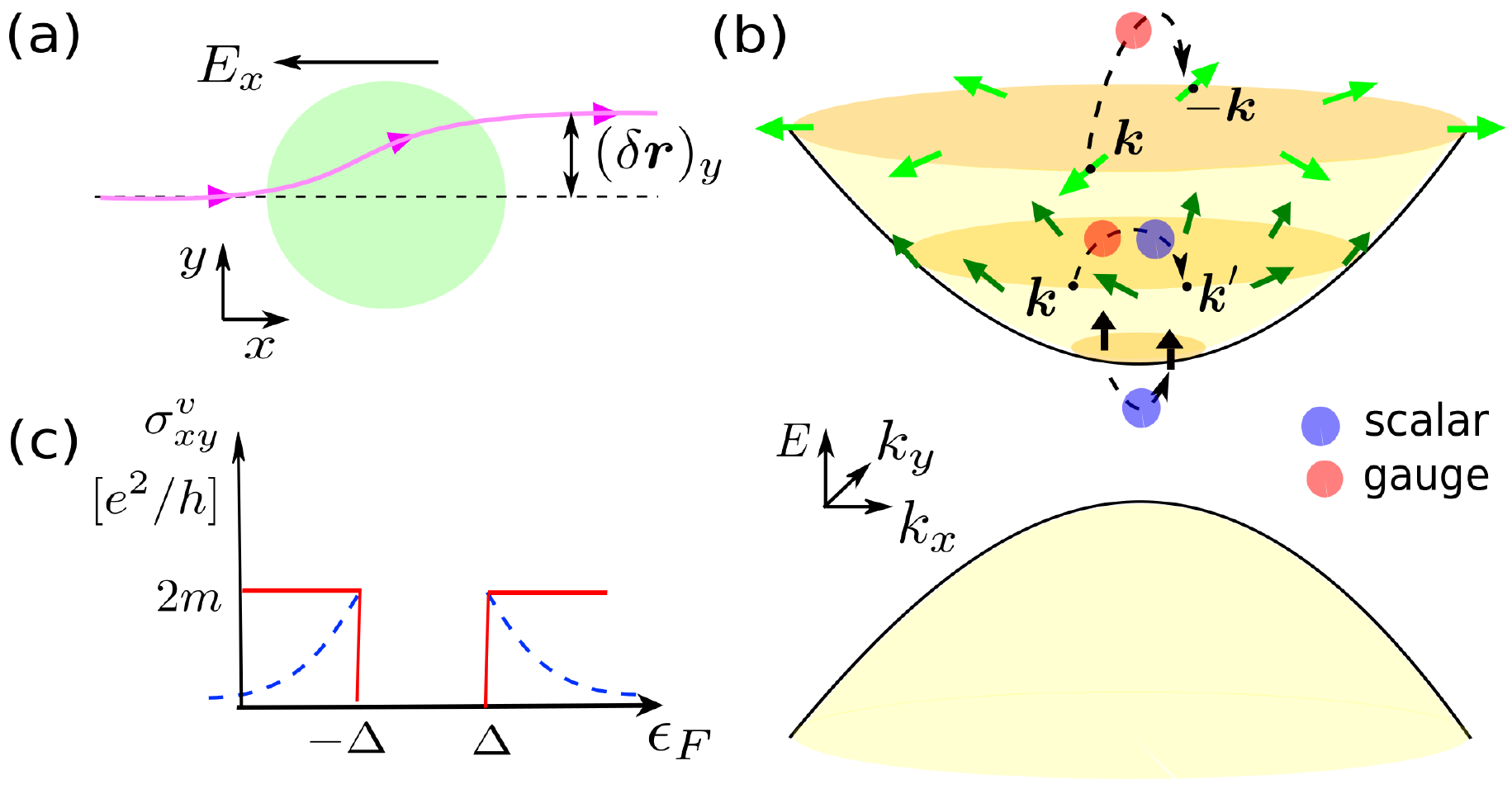}
\caption{ (a) Schematic view of coordinate shift $(\delta\bm r)_y$ for an incident wave packet accelerated by electric field $E_x$ and scattered by a gauge impurity. (b) Allowed transition process of Dirac fermions selected by different pseudospin orientation on Fermi surface and various types of disorder. Close to (Far from) the band edge, the pseudospin of electrons is aligned out of (in) the $k_x-k_y$ plane. Blue (red) dots label scalar (red) disorder. (c) Vally Hall conductivity (in units of $e^2/h$) plotted as a function of Fermi energy $\epsilon_F$ in the presence of gauge disorder. Red (blue) curve corresponds to total (side-jump) valley Hall conductivity. $2\Delta$ is the band gap and $m$ is the winding number. $m=1 (m=2)$ for monolayer (bilayer) graphene.}
\label{fig:sidejump}
\end{figure}

To reveal the effect of long-range gauge disorder on the Hall
conductivity, we invoke a recently developed semiclassical Boltzmann
theory~\cite{sinitsyn2006,sinitsyn2007}. Such theory has been widely used to investigate Hall-type transport under scalar disorder in various systems, 
whereas our work for the first time generalizes it to the long-range gauge disorder. Without loss of generality, we consider electrons scattering in the conduction band. The key quantity here is a sudden
coordinate shift~\cite{berger1970} experienced by an electron wave
packet as it scatters off an impurity (see Fig.~\ref{fig:sidejump} (a)), given by~\cite{sinitsyn2006}
\begin{equation} \label{shift}
\delta\bm r^c_{\bm k'\bm k} = \bm A^c_{\bm k'} - \bm A^c_{\bm k} - \hat{\bm D}_{\bm k',\bm k} \arg(V_{\bm k'\bm k}^c) \;,
\end{equation}
where $\bm A^c_{\bm k} = \bracket{u_{\bm k}^c|i\bm\nabla_{\bm k}u_{\bm
    k}^c}=-\sin^2\frac{\theta_{\bm k}}{2}(\hat{z}\times\bm k)/k^2$ is the Berry connection of the conduction band and $\hat{\bm D}_{\bm k',\bm k} =
\bm\nabla_{\bm k'} + \bm\nabla_{\bm k}$.  The information of the
impurity is encoded in the quantity
\begin{equation}
V^c_{\bm k'\bm k} = \frac{1}{\cal S}\int d\bm r\,
e^{-i\bm q\cdot\bm r}\bracket{u^c_{\bm k'}|V_\text{imp}(\bm r)|u^c_{\bm k}} \;,
\end{equation}
where $\cal S$ is the system area and $\bm q\equiv\bm k'-\bm k$ is the momentum transfer of electrons.  The
overall effect of the disorder is obtained by taking the disorder
average, under which the last term of Eq.~\eqref{shift} becomes
\begin{equation}
\bracket{\hat{\bm D}_{\bm k',\bm k}\arg(V_{\bm k'\bm k}^{c})}_\text{dis}
=\mathrm{Im}\frac{\bracket{V_{\bm k\bm k'}^{c}\hat{\bm D}_{\bm k',\bm k}V_{\bm k'\bm k}^{c}}_\text{dis}}{\bracket{|V_{\bm k'\bm k}^{c}|^2}_\text{dis}} \;,
\end{equation}
where $\bracket{\dots}_\text{dis}$ stands for disorder or thermal
average.
  
To proceed further, it is convenient to write $V^c_{\bm k'\bm k}$
using the chiral basis
\begin{equation}
V^{c}_{\bm k'\bm k}=d_-(\bm q)\bracket{u^c_{\bm k'}|\sigma_+|u^c_{\bm k}}
+d_+(\bm q)\bracket{u^c_{\bm k'}|\sigma_-|u^c_{\bm k}} \;,
\end{equation} 
where $\sigma_{\pm}=(\sigma_x\pm i\sigma_y)/2$. The correlation
$\bracket{|V_{\bm k'\bm k}^{c}|^2}_\text{dis}$ then breaks up into
terms with opposite chirality
$\bracket{d_{\mp}(\bm q)d_{\pm}(-\bm q)}_\text{dis}$, and terms with
the same chirality
$\bracket{d_{\pm}(\bm q)d_{\pm}(-\bm q)}_\text{dis}$.  The former contain no phase factor, whereas the latter, according to Eq.~\eqref{dq}, is proportional to $\exp(\mp 4i\varphi_{\bm q})$. Note that such term should be small after angular average, and we can neglect the term. The accuracy of this approximation is demonstrated by rigorous numerical analysis in Appendix ~\ref{sec:numerical}. By such approximation, the correlation reduces to
\begin{equation}
\begin{split}
\langle|V^{c}_{\bm k'\bm k}|^2\rangle_\text{dis}&=\frac{1}{2}\langle d_-(\bm q)d_+(\bm q)\rangle_\text{dis}\sin^2\theta_{\bm k},
\end{split} 
\end{equation}
where we have used the condition of elastic scattering, i.e., $\theta_{\bm k}=\theta_{\bm k'}$.

 This approximation also applies to the correlation
$\bracket{V_{\bm k\bm k'}^{c}\hat{\bm D}_{\bm k',\bm k}V_{\bm k'\bm k}^{c}}_\text{dis}$. Making use of the fact that
$\hat{\bm D}_{\bm k',\bm k}d_{\pm}(\bm q) = (\bm\nabla_{\bm k} +
\bm\nabla_{\bm k'})d_{\pm}(\bm k'-\bm k) = 0$, we obtain
\begin{equation}
\begin{split}\label{r_shift}
\delta\bm r^c_{\bm k'\bm k}&=\frac{\bm\Omega_c(\bm k)\times(\bm k-\bm k')}{\sin^2\theta_{\bm k}}.
\end{split}
\end{equation}
Similar calculation can be applied to electrons from valence band:
\begin{equation}
\begin{split}
\delta\bm r^v_{\bm k'\bm k}&=\frac{\bm\Omega_v(\bm k)\times(\bm k-\bm k')}{\sin^2\theta_{\bm k}}.
\end{split}
\end{equation}

We can see that in the coordinate shift $\delta\bm r^c_{\bm k'\bm k}$ or $\delta\bm r^v_{\bm k'\bm k}$, the strain-related perfactor $F(\bm q)$ drops out completely, thus this expression is generally applicable for both out-of-plane and in-plane modes
of strain fluctuations. Physically, $\delta\bm r_{\bm k'\bm k}$
describes a coordinate shift transverse to the momentum change $\bm k-\bm k'$, leading to a Hall-like current.

It is useful to compare with the short-range scalar disorder~\cite{sinitsyn2007}.  In that case, the coordinate shift is given by
\begin{equation}
\begin{split}
\delta\bm r^n_{\bm k'\bm k}&=-\frac{\bm \Omega_n(\bm k)}{|\langle u^n_{\bm k'}|u^n_{\bm k}\rangle|^2}\times(\bm k-\bm k'),
\end{split}
\end{equation}
where $n=c/v$ refers to conduction (valence) band. In the denominator $|\langle u^n_{\bm k'}|u^n_{\bm k}\rangle|^2\approx1$ near the
band edge, whereas for gauge disorder in Eq. (\ref{r_shift}) $\sin^2\theta_{\bm k}=4|\langle u^c_{\bm k'}|\sigma_+|u^c_{\bm
  k}\rangle|^2\approx0$, which makes a significant difference. Physical
meaning for this difference is that close to the band edge, pseudospin of electrons is
almost fixed, and the probability of a spin-flipping transition driven
by gauge disorder is vanishingly small (see Fig.~\ref{fig:sidejump} (b)).

Once the coordinate shift $\delta\bm r_{\bm k'\bm k}$ is derived, one
can calculate its contribution, known as the side jump, to the valley Hall conductivity. There are two different types of side-jump effects: the direct side-jump contribution $\sigma_{xy}^{direct}$ and the anomalous distribution-induced contribution $\sigma_{xy}^{adist}$~\cite{sinitsyn2006,sinitsyn2007}. We can first write down the scattering rate
\begin{equation}
\begin{split} 
\omega_{\bm k'\bm k}&=\frac{2\pi}{\hbar}\langle|V^{c}_{\bm k'\bm k}|^2\rangle_\text{dis}\delta(\epsilon_{c,\bm k}-\epsilon_{c,\bm k'})
\end{split}
\end{equation}
and the transport time
\begin{equation}
\begin{split}\label{transport}
&\frac{1}{\tau^{tr}}
=\frac{2\pi}{\hbar}\sum_{\bm k'}\langle|V^{c}_{\bm k'\bm k}|^2\rangle_\text{dis}(1-\cos(\phi_{\bm k}-\phi_{\bm k'}))\delta(\epsilon_F-\epsilon_{c,\bm k'}).
\end{split}
\end{equation}
For monolayer or bilayer graphene, point group symmetry requires that random gauge potential follows $d_{\pm}(\bm q)=F(\bm q)q^2e^{-2i\varphi_{\bm q}}$, which means $\langle d_-(\bm q)d_+(\bm q)\rangle_\text{dis}$ becomes a function of $\bm q=\bm k-\bm k'$.  Therefore the transport time $\tau^{tr}$ is isotropic for all $\bm k$ on the Fermi surface.

The coordinate shift $\delta\bm r^c_{\bm k'\bm k}$ leads to an average side-jump velocity $\bm v^{sj}(\bm k)$
\begin{equation}
\begin{split}
&v^{sj}_x(\bm k)=\sum_{\bm k'}\omega_{\bm k'\bm k}(\delta\bm r^c_{\bm k'\bm k})_x=\frac{\cos\theta_{\bm k}}{2k_F}\frac{1}{\tau^{tr}}\sin\phi_{\bm k},
\end{split}
\end{equation}
where $k_F$ is the Fermi wave vector. In the presence of an external electric field $E_y$, a nonequilibrium correction to the distribution function is given by
\begin{align}
g_{\bm k}=-\frac{\partial n_e}{\partial\epsilon_{c,\bm k}}eE_yv^c_y(\bm k)\tau^{tr}, 
\end{align}
where $n_e=1/[\exp((\epsilon_{c,\bm k}-\epsilon_F)/k_BT)+1]$ is the Fermi distribution function, and $v_y^c(\bm k)=v\sin\theta_{\bm k}\sin\phi_{\bm k}$ is the bare velocity of electrons along the electric field. As a result, at $T=0$ K, the valley Hall conductivity $\sigma_{xy}^{direct}$ (for each valley and spin) reads
\begin{equation}
\begin{split}
\sigma_{xy}^{direct}&=e\int\frac{d^2\bm k}{(2\pi)^2}\frac{g_{\bm k}}{E_y}v^{sj}_x(\bm k)=\frac{e^2}{4h}\cos\theta_{F}.
\end{split}
\end{equation}
The physical process can be understood as follows: in the weak disorder limit, the scattering rate $\omega_{\bm k'\bm k}$ is tiny, which gives rise to a small anomalous velocity $v^{sj}_x(\bm k)$ on average during the scattering events. On the other hand, weak disorder means long lifetime, i.e., electrons can be accelerated by an electric field for more time until they are stopped by disorder scattering. This creates a large correction to the Fermi distribution $g_{\bm k}$, that is, more electrons contribute to the transverse transport. As a result, a product of the small anomalous velocity and the large number of electrons lead to a disorder-independent valley Hall conductivity $\sigma_{xy}^{direct}$ as the leading-order term of the disorder potential.

In addition, $\delta\bm r^c_{\bm k'\bm k}$ can cause an anomalous distribution $g^{adist}_{\bm k}$ that also contributes to the Hall current, i.e., $\sigma_{xy}^{adist}$ term. To find its form, let us solve the equation
\begin{align}
\sum_{\bm k'}\omega_{\bm k'\bm k}(g^{adist}_{\bm k}-g^{adist}_{\bm k'}+(-\frac{\partial n_e}{\partial \epsilon_{c,\bm k}})eE_y(\delta\bm r^c_{\bm k'\bm k})_y)=0
\end{align}
to derive the nonequilibrium distribution function $g^{adist}_{\bm k}$. Take the ansatz $g^{adist}_{\bm k}=\gamma_{\bm k}k_x$, we find
\begin{equation}
\begin{split}
\gamma_{\bm k}=-(\frac{\partial n_e}{\partial \epsilon_{c,\bm k}})eE_y\frac{\cos\theta_{\bm k}}{2k^2}.
\end{split}
\end{equation}
Then at $T=0$ K, $\sigma_{xy}^{adist}$ is given by
\begin{equation}
\begin{split}
\sigma_{xy}^{adist}&=e\int\frac{d^2\bm k}{(2\pi)^2}\frac{g^{adist}_{\bm k}}{E_y}v_x^c(\bm k)=\frac{e^2}{4h}\cos\theta_{F}.
\end{split}
\end{equation}

Finally, the total side-jump valley Hall conductivity $\sigma_{xy}^{sj}$ is given by
\begin{equation}
\begin{split}\label{result}
\sigma_{xy}^{sj}&=4(\sigma_{xy}^{direct}+\sigma_{xy}^{adist})=\frac{2e^2}{h}\cos\theta_{F},
\end{split}
\end{equation}
where the factor of 4 counts the valley and spin degeneracy.  We
notice that $\sigma_{H}^\text{sj}$ reaches its maximum value at the
band edge, then decreases gradually as the Fermi energy moves away (see Fig.~\ref{fig:sidejump} (c)).
By Eq. (\ref{intrins}) and (\ref{result}), the total valley Hall conductivity then becomes
$\sigma_{H}^v=\sigma_{H}^\text{int}+\sigma_{H}^\text{sj}=2e^2/h$,
which is quantized. We have also obtained the same valley Hall conductivity by adopting a different diagrammatic approach in Appendix \ref{sec:diagram}.
Intriguingly, this gives us the same result as obtained for short-range gauge disorder using the diagrammatic approach~\cite{yang2011,yang2011a}. This indicates that valley Hall conductivity when a Dirac electron meets with gauge disorder is actually a universal quantity, which is in striking contrast to the case of scalar disorder. Note that at zero temperature our finding proposes a new type of geometric quantization of the Fermi surface, in contrast to the well-known topological quantization of the Fermi sea. This is one of the main results of our paper.

We may also apply our theory to bilayer graphene, which can be modeled
by
\begin{equation} 
H_\text{BLG} = (\hbar v)^2[(k_x^2-k_y^2)\sigma_x+2k_xk_y\sigma_y] + \Delta\sigma_z \;.
\end{equation}
After some algebra, we find
$\sigma_{H}^\text{sj}=(4e^2/h)\cos\theta_{F}$ and the total valley Hall
conductivity $\sigma_H^v = 4e^2/h$.  The doubling of $\sigma_H^v$ can
be traced back to the phase winding number of $2$ of the electrons
around the Dirac point in bilayer graphene. Note that we have neglected the weaker fluctuation of next-nearest-neighbor 
interlayer coupling in bilayer graphene~\cite{son2011}.

\section{Temperature dependence}\label{temperature}

So far we have studied the valley Hall effect at zero temperature and
its dependence on the Fermi energy.  Next we will focus on the charge
neutral point and consider the dependence of the valley Hall effect on
the band gap and temperature.  We will consider bilayer graphene systems to 
compare with the experiment~\cite{sui2015}. If the system is perfectly uniform (chemical potential $\mu=0$), $\sigma_H^v$ due to thermally-activated
carriers will be exponentially small.  For example, take
$2\Delta\sim 100$ meV in a bilayer graphene, then
$\sigma_H^v \sim 2 \times 10^{-3} e^2/h$ at $T = 70$ K, several orders
of magnitude smaller than experimental values. However, in graphene
systems the charge density typically fluctuates due to the formation
of electron-hole puddles or gate voltage fluctuation~\cite{dean2010}.
By assuming a small residue charge density $\delta n_0$, we can move
the chemical potential into the conduction or valence bands according
to
\begin{equation}
\delta n_0 =\sum_{\bm k}[n_e(\epsilon_{\bm k},\mu,T)-n_h(\epsilon_{\bm k},\mu,T)] \;,
\end{equation}
where $n_{e/h}(\epsilon_{\bm k},\mu,T)=1/[\exp(\pm(\epsilon_{c/v,\bm k}-\mu)/k_BT)+1]$ is
the Fermi distribution function for electrons (holes) and $\mu$ is the chemical potential. At finite temperatures, both electrons and holes contribute to the transport. Including all these effects, we find the extrinsic and intrinsic valley Hall conductivity
\begin{equation}
\begin{split}\label{temperature2}
&\sigma_{xy}^{direct}=\sigma_{xy}^{adist}=\frac{e^2}{4h}\frac{\Delta}{k_BT}\\
&\times\int^{\infty}_{\Delta/k_BT}dx[\frac{e^{x-\frac{\mu}{k_BT}}}{(1+e^{x-\frac{\mu}{k_BT}})^2}\frac{1}{x}+\frac{e^{x+\frac{\mu}{k_BT}}}{(1+e^{x+\frac{\mu}{k_BT}})^2}\frac{1}{x}],\\
&\sigma_{xy}^{int}=\frac{e^2}{2h}\frac{\Delta}{k_BT}\\
&\times\int^{\infty}_{\Delta/k_BT}dx[\frac{1}{1+e^{x-\frac{\mu}{k_BT}}}\frac{1}{x^2}+\frac{1}{1+e^{x+\frac{\mu}{k_BT}}}\frac{1}{x^2}],
\end{split}
\end{equation}
where the two terms in the square brackets originate from electrons and holes, respectively. For monolayer graphene, we obtain a much enhanced $\sigma_H^v$:
\begin{equation}
\begin{split}
&\sigma_{H}^{v}=4[\sigma_{xy}^{direct}+\sigma_{xy}^{adist}+\sigma_{xy}^{int}]\\
&=\frac{2e^2}{h}[\frac{1}{1+e^{(\Delta-\mu)/k_BT}}+\frac{1}{1+e^{(\Delta+\mu)/k_BT}}].
\end{split}
\end{equation}
For bilayer graphene, we will acquire an extra factor 2, as compared to monolayer case.

\begin{figure}[t]
\centering \includegraphics[width=0.38\textwidth]{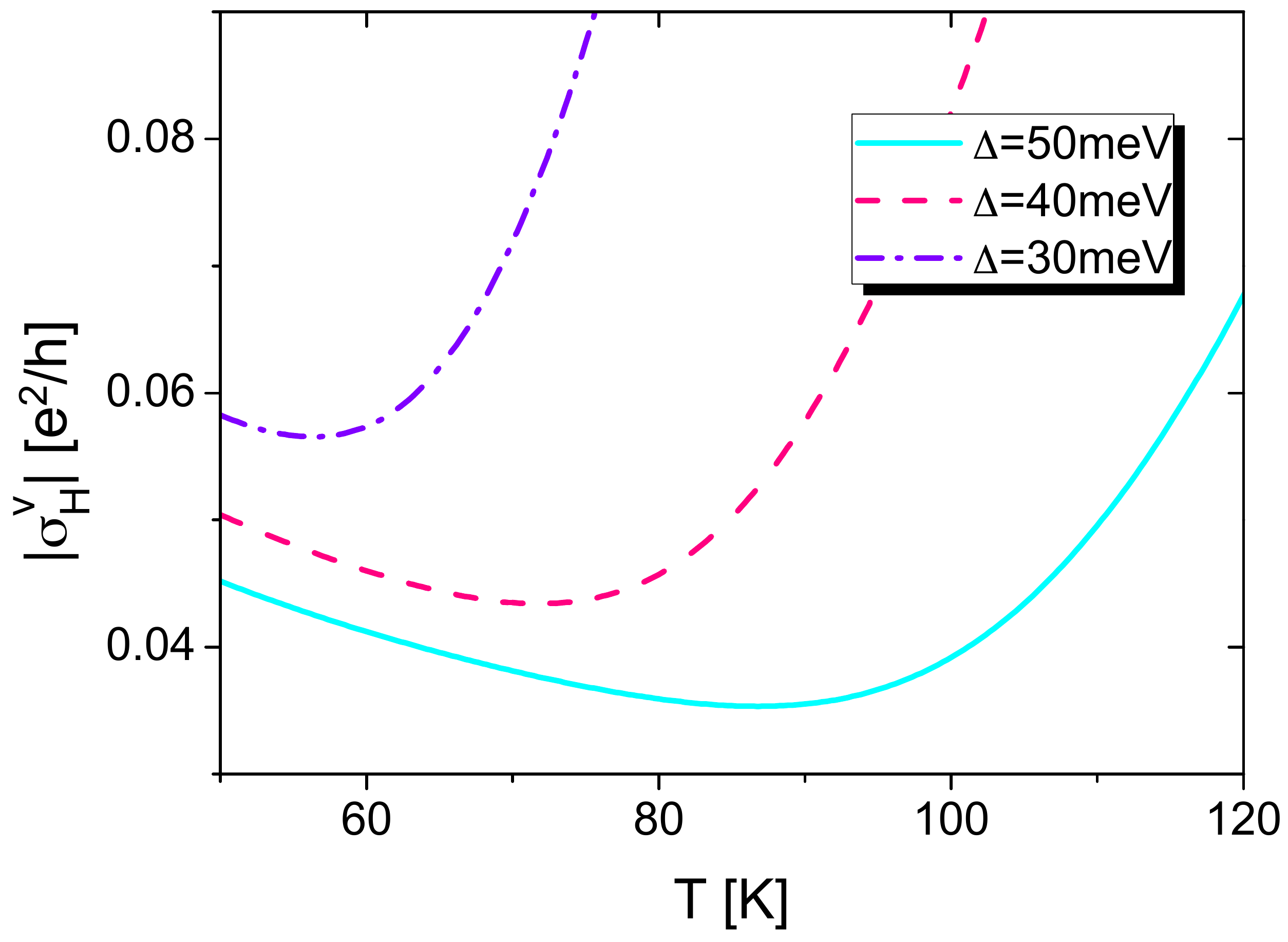}
\caption{ Temperature dependence of valley Hall conductivity $|\sigma_{H}^v|$ (in units of $e^2/h$) of bilayer graphene for $\Delta=50$, $40$, $30$meV, respectively. Parameters: $\delta n_0=1.0\times10^{10}$cm$^{-2}$.}
\label{fig:temperature}
\end{figure}

Figure~\ref{fig:temperature} shows
the calculated $\sigma_H^v$ as a function of temperature for
$\delta n_0\sim1.0\times10^{10}$ cm$^{-2}$, a value well below the
density resolution of the nonlocal peak in the
experiment~\cite{gorbachev2014}. The downwards trend at medium
temperature range $T\sim60-80$ K reproduces precisely the temperature
dependence in bilayer graphene reported in one of the
experiments (see Supplementary Figure 9b of reference ~\cite{sui2015}). Another trend found in the
experiment~\cite{sui2015}, i.e., smaller $\sigma_H^v$ with larger gap,
is also reproduced.  At higher temperature we find that $\sigma_H^v$
will increase again, although there is no available experimental data
to compare with.  As far as we know, there are no counterparts of
such peculiar temperature behaviors reported in anomalous Hall physics~\cite{nagaosa2010,xiao2018}. 
The origin of such non-monotonic temperature
behavior is the competition between intrinsic and side-jump
contributions under thermal activation.  The intrinsic contribution,
as a summation of Berry curvature over occupied states, favors
high-temperature regime with more occupied states, whereas the
side-jump contribution, suffering from a suppression of thermally
activated carriers by $1/k_BT$, favors low-temperature
regime, as shown in Eq. (\ref{temperature2}). Nevertheless, neither a gap nor temperature dependence of valley Hall conductivity has been 
discussed by previous theoretical papers~\cite{lensky2015,kirczenow2015,li2011,zhu2017,song2018,brown2018}, and our work represents a first step 
towards understanding these peculiar behaviors. Moreover, the newly predicted non-monotonic temperature
dependence can be used as a test for our theory.

\begin{figure}[t]
\centering \includegraphics[width=0.45\textwidth]{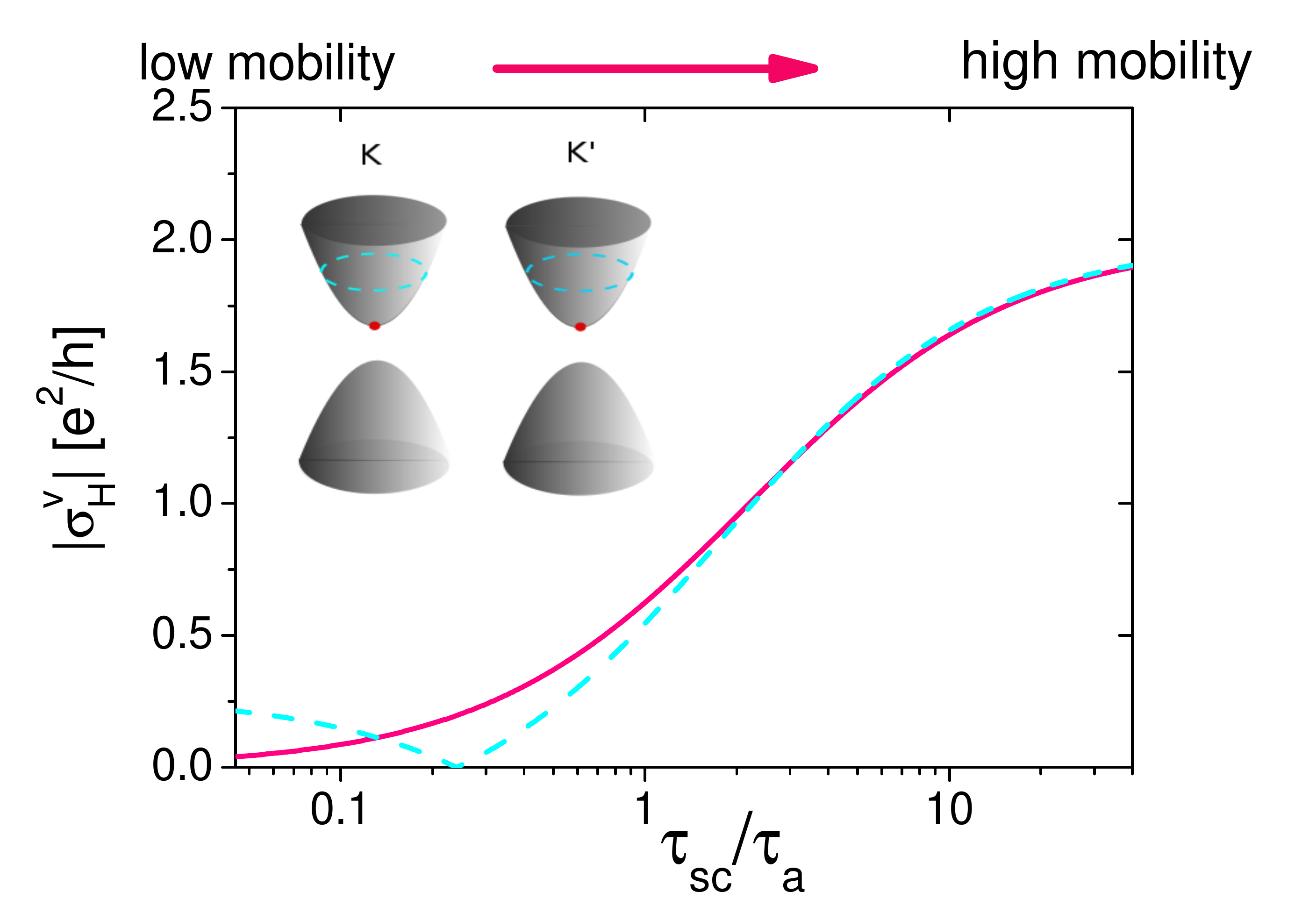}
\caption{ Magnitude of valley Hall conductivity $|\sigma_{H}^v|$ (in
  units of $e^2/h$) as functions of ratio $\tau_{sc}/\tau_a$ at $T=0$
  K. $\tau_{sc}$, $\tau_a$ refer to relaxation time for scalar and
  gauge disorder scattering, respectively. Larger $\tau_{sc}/\tau_a$
  implies more gauge disorder (high mobility sample); smaller implies more
  scalar charge disorder (low mobility sample). Red solid (cyan
  dashed) curve corresponds to doping density $n=0$ ($n=10^{10}$
  cm$^{-2}$), whose Fermi surface is indicated in the inset. Parameters
  are adopted for monolayer graphene: $\Delta=20$
  meV~\cite{gorbachev2014}, and $v=10^6$ m/s. }
\label{fig:valleyhall}
\end{figure}

\section{Numerical analysis}\label{numerics}

A crucial approximation we made in the semiclassical approach is
assuming that
$ \bracket{d_{\pm}(\bm q)d_{\pm}(-\bm q)}_\text{dis} \propto \exp(\mp
4i\varphi_{\bm q})$ is negligible after angular average.  The accuracy
of this approximation needs to be examined since it introduces a
long-range anisotropic correlation function. The effect of such correlation function is unclear yet,
and an analytical treatment seems impractical. To this end, in the
following part we develop a new numerical approach which treats the correlation function
rigorously in Appendix \ref{sec:numerical}. We consider a particular type of random strain, i.e.,
out-of-plane corrugations with
$\bm d(\bm q)=g_1|\bm q|^{-4}\mathcal{F}(\bm
q)(q_x^2-q_y^2,-2q_xq_y)$, where
$\langle\mathcal{F}(\bm q)\mathcal{F}(-\bm q)\rangle_\text{dis}$
$\propto|\bm q|^{2}$, and $g_{1}$ quantifies the electron-phonon
coupling strength in graphene~\cite{couto2014}. We find that $\sigma_{H}^{sj}$ is very
close to $ \frac{2me^2}{h}\cos\theta_{F}$ ($m=1$ for monolayer; $m=2$
for bilayer), differing by $1.69\%$ for monolayer graphene and
$0.91\%$ for bilayer graphene.  Such small deviation from the
quantized values justifies our previous treatment of dropping the
fast-oscillating part
$\langle d_{\pm}(\bm q)d_{\pm}(-\bm q)\rangle_\text{dis}$.

\section{Effect of long-range scalar potential}\label{scalar}

Besides the gauge potential $\bm d(\bm q)$, 
a long-range scalar potential may also arise due to random strain fluctuations. For
out-of-plane corrugations~\cite{couto2014}, the scalar potential is
given by $V(\bm q)=g_2|\bm q|^{-2}\mathcal{F}(\bm q)$, where $g_{2}$
is an electron-phonon coupling parameter.  Different from the gauge
potential $\bm d(\bm q)$, which represents a pseudomagnetic field,
$V(\bm q)$ is electrostatic in nature and thus it needs to be screened
by the static dielectric function $\epsilon(\bm q)=1+V_c(\bm q)N_F$
within the Thomas-Fermi approximation~\cite{ando2006,dassarma2011},
where $N_F$ is the density of states,
$V_c(\bm q)=2\pi e^2/\epsilon_0|\bm q|$ is the two-dimensional Coulomb
potential, and $\epsilon_0$ is the background (including the
substrate) dielectric constant. Here we will focus on the strong
screening regime (see Appendix \ref{sec:screen}): $\epsilon_0\ll e^2N_F/k_F$, which
characterizes the charge neutral point of gapped graphene systems.

Taking into account both the gauge and scalar potentials, we find that
the total valley Hall conductivity of a gapped graphene at $T=0$ K is (see Appendix \ref{sec:mix}):
\begin{equation}\label{scalar_pot}
\sigma_{H}^{v} =\frac{2e^2}{h}[1-\frac{(4+\sin^2\theta_{F})\cos\theta_{F}}{\frac{4}{w_0}\frac{\tau_{sc}}{\tau_a}+4-3\sin^2\theta_{F}}].
\end{equation}
Here $w_0=\sqrt{(k_FW/\pi)^2-1}/\pi$ is a cutoff-related factor with the device width $W$. For $W=1$ $\mu$m and residue charge density
$\delta n_0=1.0\times10^{10}$ cm$^{-2}$, we estimate $w_0\approx2.2$.
The ratio $\tau_{sc}/\tau_a=(2e^2g_1/\hbar\epsilon_0g_2)^2w_0$ defines
the relative strength between the scalar and gauge disorder.  In the
limit $\tau_{sc}/\tau_a\rightarrow\infty$, i.e., in the absence of
scalar disorder, $\sigma_H^v$ reduces to
$2e^2/h$. Figure~\ref{fig:valleyhall} shows the calculated
$\sigma_H^v$ as a function of $\tau_{sc}/\tau_a$ at two different
Fermi energies.  From this figure, one can immediately understand why
nonlocal signal is only measured in high-quality graphene on hBN
rather than on SiO$_2$ substrate~\cite{zyb2015}, since in the former
(latter) case gauge (scalar) disorder is the dominant source of
disorder scattering, corresponding to the limit
$\tau_{sc}/\tau_a\rightarrow\infty$ ($\tau_{sc}/\tau_a\rightarrow0$).

\section{Conclusion and discussion}\label{conclusion}

We have provided an alternative scenario to understand 
the large valley Hall conductivity observed in experiments, based on scattering 
from random strain fluctuations. The origin is intimately related to an enhanced 
coordinate shift under gauge disorder scattering in Dirac systems. Temperature 
and gap dependence is qualitatively reproduced. Our work paves the way for studying the 
effect of classical strain modes, or phonon modes~\cite{zhang2015}, on the transverse transport of broad classes of 2D materials and van
der Waals heterostructures.

A few remarks are in order. Note that our theory is only valid for the band transport
regime; the valley Hall effect in the phonon-assisted variable-range-hopping
regime~\cite{zou2010} still remains an open question.  In addition, we
would like to point out that a direct comparison of our result to the
experiments requires a careful extraction of the valley Hall
conductivity from the nonlocal measurement~\cite{gorbachev2014,sui2015,shimazaki2015}.  In particular, an
accurate determination of the valley diffusion length $\ell_v$ is
crucial~\cite{sui2015} since the nonlocal signals depend on $\ell_v$
exponentially~\cite{abanin2009}. Then the ``smoking gun" validation of our theory, i.e., a non-monotonic 
temperature behaviors of valley Hall conductivity, can be examined. As a supplement, weak-localization
magneto-resistance measurement~\cite{couto2014} and Raman spectroscopy~\cite{neumann2015} can also be used to uncover the role of strain 
fluctuations. A complete understanding of the
valley Hall effect thus requires further experimental and theoretical
efforts.

A universal transverse Imbert-Fedorov (IF) shift of electrons was also discovered in 
Weyl semimetals~\cite{jiang2015,yang2015}. The universal coordinate shift proposed in our work differs from the IF
 shift in two ways. First, the IF shift appears in the three-dimensional massless Weyl fermions, and has no counterpart in 2D,
  while the coordinate shift is relevant to the 2D massive Dirac fermions. Second, the IF shift occurs at normal interface,
   while the coordinate shift becomes universal only when scattered by gauge disorder. Appealingly, the considerations in this work 
   can be generalized to describe other 2D systems with topological properties, such as superconductors~\cite{yu2018}, excitons~\cite{onga2017}, 
   plasmons~\cite{shili2018}, polaritons~\cite{gutierrez2018}, or under magnetic field~\cite{komatsu2018}.

Recently, experimental observations of valley Hall transport have also been made in 
atomically thin MoS2 systems~\cite{wu2018,hung2019}. The fact that monolayer and trilayer MoS2
share qualitatively similar behaviors of valley Hall signals~\cite{wu2018} indicates that the details of 
Berry curvature distribution in the conduction bands may have little influence on the final result.
This is actually consistent with our theoretical prediction of the article.

\section*{ Acknowledgements}

We are grateful to Qian Niu, Shengyuan A. Yang, Mengqiao Sui, and Yuanbo Zhang for
stimulating discussions.  This work is supported by DOE BES Pro-QM EFRC (DE-SC0019443). W. Y. S. also acknowledges the support of 
a startup grant from Guangzhou University.

\appendix

\section{Diagrammatic approach to mixed gauge and scalar disorder}\label{sec:diagram}

In this section we present a full quantum mechanical treatment using the diagrammatic approach~\cite{sinitsyn2007,shan2013} to study the valley Hall effect in the presence of out-of-plane corrugations~\cite{couto2014}. The purpose of this section is two fold.  The diagrammatic approach provides an additional check of the semiclassical result.  In addition, this approach is systematic, and can be applied to mixed gauge and scalar disorder.  It is also convenient for numerical calculations when we consider the anisotropic long-range correlation functions.  In the following we will focus on monolayer graphene. 

\subsection{Correlation function}

For out-of-plane corrugation the induced vector potential $\bm d(\bm q)$ and scalar potential $V(\bm q)$ are given by~\cite{couto2014}
\begin{equation}
\begin{split}\label{strain}
\bm d(\bm q)&=g_1\frac{1}{|\bm q|^4}\mathcal{F}(\bm q)(q_x^2-q_y^2,-2q_xq_y) \;, \\
V(\bm q)&=g_2\frac{1}{|\bm q|^2}\mathcal{F}(\bm q) \;,
\end{split}
\end{equation}
where 
\begin{align}\label{strain2}
\mathcal{F}(\bm q)&=-\int d\bm q_1h(\bm q_1)h(\bm q-\bm q_1)(\bm q\times\bm q_1)^2 \;.
\end{align}
$h(\bm q)$ is the Fourier transform of the height field $h(\bm r)$, and $g_{1,2}$ quantify the electron-phonon coupling strength in graphene. We assume the height correlation is $\langle h(\bm q)h(-\bm q)\rangle_\text{dis}\propto|\bm q|^{-4}$, from which one finds $\langle\mathcal{F}(\bm q)\mathcal{F}(-\bm q)\rangle_\text{dis}=C|\bm q|^2$, where $C$ is a material-dependent parameter. The Born scattering amplitude is given by
\begin{equation}
\begin{split}
U_{cc}^A(\bm q)&=\int\frac{d\bm r}{S}e^{-i\bm q\cdot\bm r}\langle u_{\bm k'}^c|\bm d(\bm r)\cdot\bm\sigma|u_{\bm k}^c\rangle\\
&=\frac{\sin\theta_{\bm k}}{2S}[d_-(\bm q)e^{i\phi_{\bm k}}+d_+(\bm q)e^{-i\phi_{\bm k'}}],\\
U_{cc}^V(\bm q)&=\int\frac{d\bm r}{S}e^{-i\bm q\cdot\bm r}\langle u^c_{\bm k'}|V(\bm r)|u^c_{\bm k}\rangle\\
&=\frac{V(\bm q)}{S}[\cos^2\frac{\theta_{\bm k}}{2}+\sin^2\frac{\theta_{\bm k}}{2}e^{i(\phi_{\bm k}-\phi_{\bm k'})}],
\end{split}
\end{equation}
where $d_{\pm}(\bm q)=g_1q^2_{\mp}\mathcal{F}(\bm q)/|\bm q|^4$ and $V(\bm q)=g_2\mathcal{F}(\bm q)/|\bm q|^2$. The identity $\bm q=\bm k'-\bm k$ leads to a useful relation $q_{\pm}=k(e^{\pm i\phi_{\bm k'}}-e^{\pm i\phi_{\bm k}})$. The correlation functions are
\begin{align}
&\langle U_{cc}^A(\bm q)U_{cc}^A(-\bm q)\rangle_\text{dis}=\frac{g_1^2\sin^2\theta_{\bm k}}{4S^2}\langle\mathcal{F}(\bm q)\mathcal{F}(-\bm q)\rangle_\text{dis}\nonumber\\
&\times[\frac{q_+^4}{|\bm q|^8}e^{i(\phi_{\bm k}+\phi_{\bm k'})}+\frac{q_-^4}{|\bm q|^8}e^{-i(\phi_{\bm k}+\phi_{\bm k'})}+\frac{2}{|\bm q|^4}], \label{correlation1} \\
&\langle U_{cc}^V(\bm q)U_{cc}^V(-\bm q)\rangle_\text{dis}=\frac{g_2^2}{|\bm q|^4S^2}\langle\mathcal{F}(\bm q)\mathcal{F}(-\bm q)\rangle_\text{dis}\nonumber\\
&\times(\cos^4\frac{\theta_{\bm k}}{2}+\sin^4\frac{\theta_{\bm k}}{2}+2\cos^2\frac{\theta_{\bm k}}{2}\sin^2\frac{\theta_{\bm k}}{2}\cos(\phi_{\bm k}-\phi_{\bm k'})).\label{correlation2}
\end{align}
for gauge and scalar disorder, respectively.  Similar to what we did in the main text, we will ignore the first two terms in the square bracket on the right-hand side of Eq. (\ref{correlation1}) and (\ref{correlation2}). These terms originate from $\langle d_{\pm}(\bm q)d_{\pm}(\bm q)\rangle_\text{dis}$, whose contributions are vanishingly small and can be neglected. The validity of this approximation will be demonstrated in the next section. Under this approximation, the correlation function for gauge disorder becomes
\begin{align}
\langle U_{cc}^A(\bm q)U_{cc}^A(-\bm q)\rangle_\text{dis}&=\frac{g_1^2\sin^2\theta_{\bm k}}{2S^2|\bm q|^4}\langle\mathcal{F}(\bm q)\mathcal{F}(-\bm q)\rangle_\text{dis}.\label{correlation3}
\end{align}

\subsection{Relaxation time and longitudinal conductivity}\label{sec:screen}

We can use the correlation function in Eq. (\ref{correlation2}) and (\ref{correlation3}) to derive the relaxation time. According to Fermi's golden rule, the relaxation time for gauge disorder reads
\begin{equation}
\begin{split}
\frac{1}{\tau_A}&=\frac{2\pi}{\hbar}\sum_{\bm k'}\langle U_{cc}^A(\bm q)U_{cc}^A(-\bm q)\rangle_\text{dis}\delta(\epsilon_F-\epsilon_{c,\bm k'})\\
&=\frac{\pi N_FCg_1^2}{2S\hbar k_F^2}w_0\sin^2\theta_{\bm k},\label{tau}
\end{split}
\end{equation}
where $N_F=\epsilon_F/2\pi\hbar^2v^2$ is the density of states per spin and valley. The cutoff factor $w_0$, given by
\begin{align}\label{omega0}
w_0&=\int_{\phi_0}^{2\pi-\phi_0}\frac{d\phi_{\bm k}}{2\pi}\frac{1}{1-\cos\phi_{\bm k}}=\frac{\cot\frac{\phi_0}{2}}{\pi} \;,
\end{align}
is introduced to remove the divergence at small momentum~\cite{vozmediano2010}, whose physical origin is due to the finite-size effect of samples. Consider a sample with width $W$, then $|\bm q|\geq q_0=\frac{\pi}{W}$. This corresponds to a cutoff angle $\phi_0=2\arcsin(q_0/2k_F)$, and hence $w_0=\sqrt{(k_FW/\pi)^2-1}/\pi$.

On the other hand, for scalar potential, we need to take into account the screening effect. Based on the Thomas-Fermi approximation~\cite{ando2006,dassarma2011}, we can replace $U^V(\bm q)$ by $U^V(\bm q)/\epsilon(\bm q)$, where $\epsilon(\bm q)$ is the dielectric function, $\epsilon(\bm q)=1+g_sg_vV_c(\bm q)N_F$, $V_c(\bm q)=2\pi e^2/\epsilon_0|\bm q|$ is the two-dimensional Coulomb potential, $\epsilon_0$ is the background (including the substrate) dielectric constant, and $g_s=2$ $(g_v=2)$ refers to the spin (valley) degeneracy. Now we can write down the relaxation time
\begin{equation}
\begin{split}
&\frac{1}{\tau_V}=\frac{2\pi}{\hbar}\sum_{\bm k'}\frac{\langle U_{cc}^V(\bm q)U_{cc}^V(-\bm q)\rangle_\text{dis}}{\epsilon^2(\bm q)}\delta(\epsilon_F-\epsilon_{c,\bm k'}) \\
&=\frac{2\pi g_2^2C}{\hbar S}N_F\int\frac{d\varphi_{\bm k'}}{2\pi}\frac{\cos^4\frac{\theta_{\bm k}}{2}+\sin^4\frac{\theta_{\bm k}}{2}+2\cos^2\frac{\theta_{\bm k}}{2}\sin^2\frac{\theta_{\bm k}}{2}\cos\varphi_{\bm k'}}{(\sqrt{2}k_F\sqrt{1-\cos\varphi_{\bm k'}}+\frac{8\pi e^2N_F}{\epsilon_0})^2} \\
&\approx\frac{\epsilon_0^2g_2^2C}{32\pi\hbar Se^4N_F}(1-\frac{1}{2}\sin^2\theta_{\bm k}),\   \  \epsilon_0\ll e^2N_F/k_F.
\end{split}
\end{equation}
In the last step of above derivation, we have taken the strong-screening limit: $\epsilon_0\ll e^2N_F/k_F$. The physics around the charge neutral point belongs to this limit, within which the correlation effectively becomes a short-range one.

\begin{figure}[t]
\centering \includegraphics[width=0.40\textwidth]{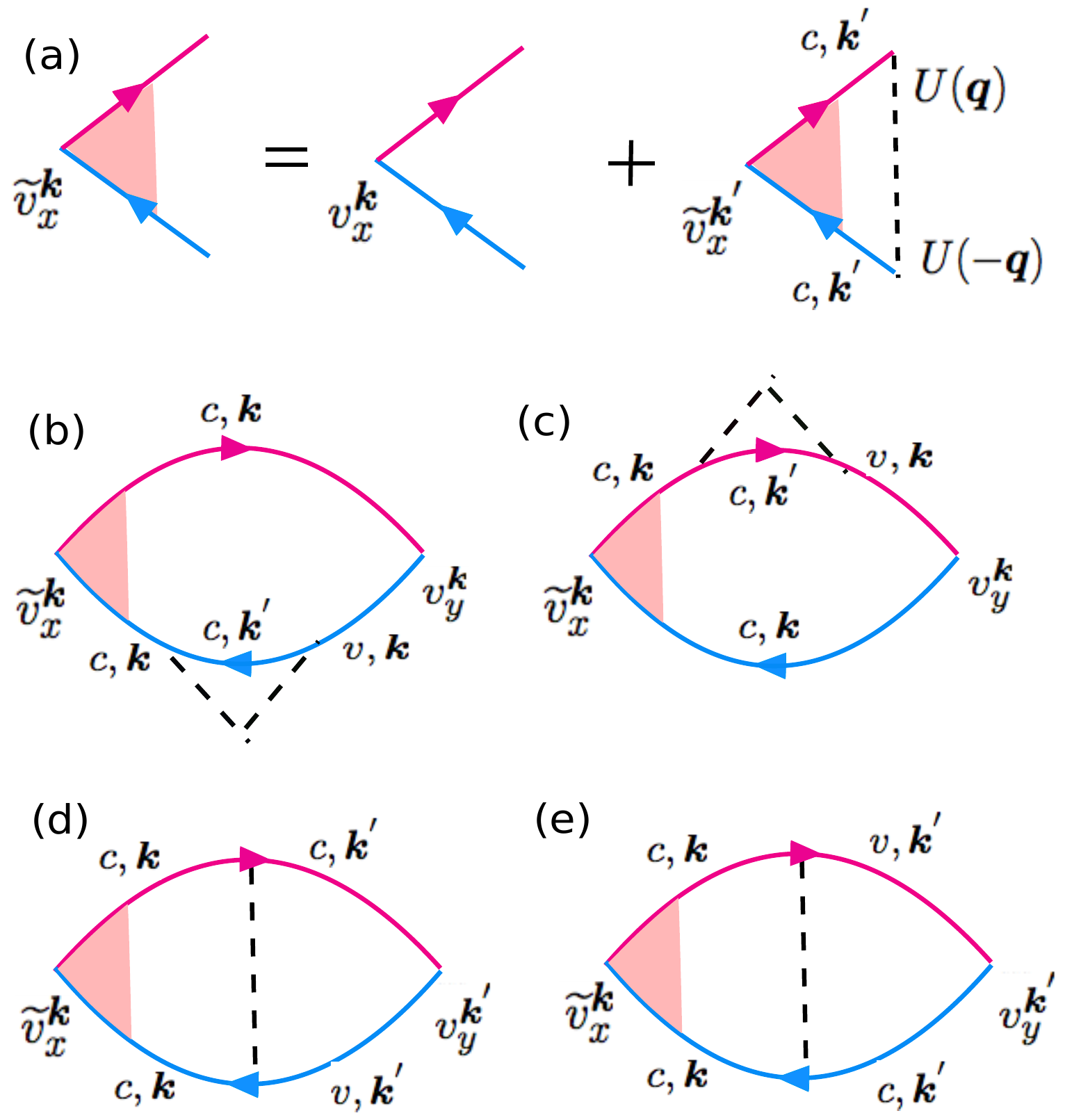}
\caption{ (a) Ladder diagram correction to the velocity vertex. (b)-(e) Diagrams corresponding to the side jump contribution to the Hall conductivity.
$\tilde{v}_x^{\bm k}$ ($v_x^{\bm k}$) refers to the modified (unmodified) velocity. Red (blue) solid line corresponds to the retarded (advanced) Green's function and dashed line represents the disorder-averaged correlation function.}
\label{fig:diagram}
\end{figure}

Next we need to figure out the modified velocity $(\tilde{v}_{x}^{\bm k})_{cc}$ due to the intraband vertex correction. According to Fig \ref{fig:diagram}, we have
\begin{align}
(\tilde{v}_{x}^{\bm k})_{cc}&=(v_x^{\bm k})_{cc}\nonumber\\
&+\sum_{\bm k'}\langle U_{cc}^{A}(\bm q)U_{cc}^{A}(-\bm q)\rangle_\text{dis} G^r_{c,\bm k'}G^a_{c,\bm k'}(\tilde{v}_{x}^{\bm k'})_{cc} \label{vertex1}
\end{align}
and
\begin{align}
(\tilde{v}_{x}^{\bm k})_{cc}&=(v_x^{\bm k})_{cc}\nonumber\\
&+\sum_{\bm k'}\frac{\langle U_{cc}^{V}(\bm q)U_{cc}^{V}(-\bm q)\rangle_\text{dis}}{\epsilon^2(\bm q)}G^r_{c,\bm k'}G^a_{c,\bm k'}(\tilde{v}_{x}^{\bm k'})_{cc}\label{vertex2}
\end{align}
for gauge and scalar disorder, respectively. Here the bare velocity is given by $(v_x^{\bm k})_{cc}\equiv\langle u^c_{\bm k}|\hat{v}_x|u^c_{\bm k}\rangle=v\sin\theta_{\bm k}\cos\varphi_{\bm k}$, and retarded (advanced) Green's function is $G_{c,\bm k}^{r/a}=1/(\epsilon_F-\epsilon_{c,\bm k}\pm i\hbar/2\tau_j)$, with $j=A,V$. To solve the equations, we take the ansatz $(\tilde{v}_{x}^{\bm k})_{cc}=\eta(v_{x}^{\bm k})_{cc}$ and substitute it into Eqs. (\ref{vertex1}) and (\ref{vertex2}).  We find $\eta=w_0$ for gauge disorder and $\eta=\frac{4(1-\frac{1}{2}\sin^2\theta_{\bm k})}{1+3\cos^2\theta_{\bm k}}$ for scalar disorder, respectively. Based on this, we can obtain the transport time
\begin{equation}
\begin{split}
\tau^{tr}_A&\equiv\eta\tau_A=\frac{4S\hbar \epsilon_F}{Cg_1^2},\\
\tau^{tr}_V&\equiv\eta\tau_V\approx\frac{128\pi\hbar Se^4N_F}{\epsilon_0^2g_2^2C}\frac{1}{1+3\cos^2\theta_{\bm k}},
\end{split}
\end{equation}
which agree with previous results by using the Boltzmann approach~\cite{couto2014}. 

\subsection{Valley Hall conductivity}\label{sec:mix}

For scalar disorder in the strong-screening limit $\epsilon_0\ll e^2N_F/k_F$, the correlation function $\langle U^V(\bm q)U^V(-\bm q)\rangle_\text{dis}/\epsilon^2(\bm q)$ exhibits a short-range behavior, whose extrinsic valley Hall conductivity has been obtained before~\cite{sinitsyn2007,yang2011}:
\begin{equation}
\begin{split}
\sigma_{xy}=-\frac{2e^2}{h}\frac{\sin^2\theta_{\bm k}\cos\theta_{\bm k}}{1+3\cos^2\theta_{\bm k}}
\end{split}
\end{equation}
for each spin and valley. In this expression, higher-order side-jump or skew-scattering contributions have been ignored since we are interested in the low doping regime. Next we turn to the gauge disorder, which has not been studied before. Fig.~\ref{fig:diagram} (b)-(e) correspond to the leading-order side-jump contributions to the valley Hall conductivity in the weak scattering limit, given by  
\begin{equation}
\begin{split}\nonumber
\sigma_{xy}^{b,i}&=\frac{e^2\hbar}{2\pi S}\sum_{\bm k,\bm k'}(\tilde{v}_{x}^{\bm k})_{cc}(v_{y}^{\bm k})_{cv}\\
&\times G^r_{c,\bm k}G^a_{c,\bm k}G^a_{v,\bm k}G^a_{c,\bm k'}\langle U_{vc}^A(-\bm q)U_{cc}^A(\bm q)\rangle_\text{dis},\\
\sigma_{xy}^{c,i}&=\frac{e^2\hbar}{2\pi S}\sum_{\bm k,\bm k'}(\tilde{v}_{x}^{\bm k})_{cc}(v_{y}^{\bm k})_{vc}\\
&\times G^r_{c,\bm k}G^r_{v,\bm k}G^r_{c,\bm k'}G^a_{c,\bm k}
\langle U_{cv}^A(\bm q)U_{cc}^A(-\bm q)\rangle_\text{dis},\\
\sigma_{xy}^{d,i}&=\frac{e^2\hbar}{2\pi S}\sum_{\bm k,\bm k'}(\tilde{v}_{x}^{\bm k})_{cc}(v_{y}^{\bm k'})_{cv}\\
&\times G^r_{c,\bm k}G^a_{c,\bm k}G^r_{c,\bm k'}G^a_{v,\bm k'}\langle U_{vc}^A(\bm q)U_{cc}^A(-\bm q)\rangle_\text{dis},
\end{split}
\end{equation}

\begin{equation}
\begin{split}\nonumber
\sigma_{xy}^{e,i}&=\frac{e^2\hbar}{2\pi S}\sum_{\bm k,\bm k'}(\tilde{v}_{x}^{\bm k})_{cc}(v_{y}^{\bm k'})_{vc}\\
&\times G^r_{c,\bm k}G^a_{c,\bm k}G^r_{v,\bm k'}G^a_{c,\bm k'}\langle U_{cv}^A(-\bm q)U_{cc}^A(\bm q)\rangle_\text{dis}.
\end{split}
\end{equation}
By making use of $[U^i_{nn'}(\bm q)]^*=U^i_{n'n}(-\bm q)$, $i=A,V$, $n,n'=c,v$, we find the following symmetry properties:
\begin{align}
\sigma_{xy}^{c,i}&=[\sigma_{xy}^{b,i}]^*,\   \ \sigma_{xy}^{e,i}=[\sigma_{xy}^{d,i}]^*.
\end{align}
After some algebra, we find that for gauge disorder
\begin{equation}
\begin{split}
\sigma_{xy}^{b,A}+\sigma_{xy}^{c,A}&=\frac{e^2}{4h}w_0\cos\theta_{\bm k}, \\
\sigma_{xy}^{d,A}+\sigma_{xy}^{e,A}&=\frac{e^2}{4h}(1-w_0)\cos\theta_{\bm k},
\end{split}
\end{equation}
and thus
\begin{align}
\sigma_{xy}^{b,A}+\sigma_{xy}^{c,A}+\sigma_{xy}^{d,A}+\sigma_{xy}^{e,A}&=\frac{e^2}{4h}\cos\theta_{\bm k}.
\end{align}
Moreover, there are equivalent contributions from diagrams by rotating Fig. \ref{fig:diagram} (b)-(e) by $180^{\circ}$, then exchanging the subscript $x$, $y$~\cite{sinitsyn2007}. Therefore the total extrinsic valley Hall conductivity (including spin degeneracy) reads $\frac{2e^2}{h}\cos\theta_{\bm k}$, which reproduces Eq. (\ref{result}) by applying the semiclassical approach.

Furthermore we can study the situation with mixed gauge and scalar disorder. In this case, the total relaxation time is given by
\begin{equation}
\begin{split}
\frac{1}{\tau}&=\frac{1}{\tau_A}+\frac{1}{\tau_V}.
\end{split}
\end{equation}
For convenience, we can introduce a Fermi-energy-independent relaxation time $\tau_a$, $\tau_{sc}$ for gauge and scalar disorder, respectively: 
\begin{equation}
\begin{split}
\tau_A&\equiv\frac{\tau_a}{\cos\theta_{\bm k}},\  \  \tau_a=\frac{4S\hbar\Delta}{Cg_1^2w_0}, \\
\tau_V&\equiv\frac{\tau_{sc}}{\cos\theta_{\bm k}(1-\frac{1}{2}\sin^2\theta_{\bm k})},\   \ \tau_{sc}=\frac{16Se^4\Delta}{\epsilon_0^2g_2^2C\hbar v^2}.
\end{split}
\end{equation}
For such mixed disorder, we can follow the same procedure as above and derive
\begin{equation}
\begin{split}
&\sigma_{xy}^{b}+\sigma_{xy}^{c}=-\frac{e^2}{4h}\eta\cos^2\theta_{\bm k}[-\frac{\tau}{\tau_a}+\frac{\tau}{2\tau_{sc}}\sin^2\theta_{\bm k}],\\
&\sigma_{xy}^{d}+\sigma_{xy}^{e}=-\frac{e^2}{4h}\eta\cos^2\theta_{\bm k}
[\frac{\tau}{\tau_a}(1-\frac{1}{w_0})+\frac{\tau}{2\tau_{sc}}\sin^2\theta_{\bm k}],\\
&\sigma_{xy}^{b}+\sigma_{xy}^{c}+\sigma_{xy}^{d}+\sigma_{xy}^{e}\\
&=-\frac{e^2}{4h}\eta\cos^2\theta_{\bm k}
[-\frac{\tau}{\tau_a}\frac{1}{w_0}+\frac{\tau}{\tau_{sc}}\sin^2\theta_{\bm k}],
\end{split}
\end{equation}
where the correction factor
\begin{align}
\eta&=\frac{1}{1-\frac{\tau}{\tau_a}(1-\frac{1}{w_0})\cos\theta_{\bm k}-\frac{\tau}{4\tau_{sc}}\cos\theta_{\bm k}\sin^2\theta_{\bm k}}.
\end{align}
Finally, by adding the intrinsic term, we find the total contribution is
\begin{align}
\sigma_{xy}^{v}&=\frac{2e^2}{h}[1-\frac{(4+\sin^2\theta_{\bm k})\cos\theta_{\bm k}}{\frac{4}{w_0}\frac{\tau_{sc}}{\tau_a}+4-3\sin^2\theta_{\bm k}}],
\end{align}
which gives Eq. (\ref{scalar_pot}) in the main text.

\section{\label{sec:numerical}Numerical treatment of anisotropic correlation function}

In our derivation, we have made use of the approximation that terms $\langle d_{\pm}(\bm q)d_{\pm}(\bm q)\rangle_\text{dis}$ in the correlation function vanish after angular average.  In this section, we test this approximation by treating the correlation function exactly, i.e., keeping all the terms in Eq. (\ref{correlation1}) and (\ref{correlation2}). We focus on a generic type of gauge disorder satisfying $\langle\mathcal{F}(\bm q)\mathcal{F}(-\bm q)\rangle_\text{dis}=C|\bm q|^{2\epsilon+2}$, where the value of $\epsilon$ depends on the microscopic details. For example, $\epsilon=0$ and $1$ corresponds to thermally excited and substrate-induced ripples, respectively~\cite{vozmediano2010}. We also consider a generic chiral model
\begin{align}
H=\begin{pmatrix}
\Delta & Ak_-^m \\
Ak_+^m & -\Delta \\
\end{pmatrix},
\end{align}
where $m=1$ and $2$ correspond to monolayer and bilayer graphene, respectively. This leads to the eigenvalue and eigenstates
\begin{equation}
\begin{split}
\epsilon_{c/v,\bm k}&=\pm\sqrt{\Delta^2+A^2k^{2m}},\\
|u^c_{\bm k}\rangle&=\left(\begin{array}{cc}
\cos\frac{\theta_{\bm k}}{2} \\
\sin\frac{\theta_{\bm k}}{2}e^{im\phi_{\bm k}} \\
\end{array}\right),\    \  |u^v_{\bm k}\rangle=\left(\begin{array}{cc}
\sin\frac{\theta_{\bm k}}{2} \\
-\cos\frac{\theta_{\bm k}}{2}e^{im\phi_{\bm k}} \\
\end{array}\right),
\end{split}
\end{equation}
where $\cos\theta_{\bm k}=\Delta/\epsilon_{c,\bm k}$, $\sin\theta_{\bm k}=Ak^m/\epsilon_{c,\bm k}$.  The density of states is given by $N_F=\epsilon_F/(2\pi mA^2k_F^{2m-2})$. We can write down a complete form of the correlation function
\begin{equation}
\begin{split}
&\langle U_{cc}^A(\bm q)U_{cc}^A(-\bm q)\rangle_\text{dis}=\frac{Cg_1^2}{4S^2|\bm q|^{6-2\epsilon}}\sin^2\theta_{\bm k}\\
&\times[k^4e^{im(\phi_{k'}-\phi_k)}e^{i(4+2m)\phi_{\bm k}}(e^{i(\phi_{\bm k'}-\phi_k)}-1)^4\\
&+k^4e^{-im(\phi_{k'}-\phi_k)}e^{-i(4+2m)\phi_{\bm k}}(e^{-i(\phi_{\bm k'}-\phi_k)}-1)^4+2q^4].
\end{split}
\end{equation}
For monolayer graphene with thermal ripples: $m=1$, $\epsilon=0$, it reduces to Eq. (\ref{correlation1}). Similar to Eq. (\ref{tau}), we can evaluate the relaxation time by
\begin{equation}
\begin{split}
\frac{1}{\tau_{\bm k}}&=\frac{\pi N_FCg_1^2}{2^{1-\epsilon}S\hbar k^{2-2\epsilon}}\sin^2\theta_{\bm k}[w_{m+2}\cos(4+2m)\phi_{\bm k}+w_0],
\end{split}
\end{equation}
where a cutoff factor is introduced
\begin{align}\label{omegam}
w_{m}&=\int_{\phi_0}^{2\pi-\phi_0}\frac{d\phi_{\bm k}}{2\pi}\frac{\cos m\phi_{\bm k}}{(1-\cos\phi_{\bm k})^{1-\epsilon}}.
\end{align}
For such anisotropic problem, it is convenient to establish a self-consistent equation for the mean free path $(L_{x}^{\bm k})_{cc}$~\cite{Tokura1998}
\begin{equation}
\begin{split}\label{self}
\frac{1}{\tau_{\bm k}}(L_{x}^{\bm k})_{cc}&=(v_x^{\bm k})_{cc}\\
&+\frac{2\pi N_FS}{\hbar}\int\frac{d\phi_{\bm k'}}{2\pi}\langle U_{cc}^{A}(\bm q)U_{cc}^{A}(-\bm q)\rangle_\text{dis}(L_{x}^{\bm k'})_{cc}.
\end{split}
\end{equation}
Since the bare velocity follows $(v_x^{\bm k})_{cc}=\frac{mA}{\hbar}k^{m-1}\sin\theta_{\bm k}\cos\phi_{\bm k}$, we can take the following ansatz,
\begin{align}
(L_{x}^{\bm k})_{cc}&=\frac{2^{1-\epsilon}k^{1-2\epsilon+m}S}{\pi N_FCg_1^2\sin\theta_{\bm k}}mA(\sum_{n=0}^{\infty}f_n\cos n\phi_{\bm k}).
\end{align}
By substituting it into Eq. (\ref{self}), we find that the coefficients $f_n$ satisfy
\begin{align}
\left(\begin{array}{cccccc}
f_1 \\
f_{2m+3} \\
f_{2m+5} \\
f_{4m+7} \\
f_{4m+9} \\
\vdots
\end{array}\right)=T^{-1}
\left(\begin{array}{cccccc}
2 \\
0 \\
0 \\
0 \\
0 \\
\vdots
\end{array}\right),
\end{align}
where the matrix $T$ has a non-closed form
\begin{widetext}
\begin{eqnarray}
T=\begin{pmatrix}
2(w_0-w_1) & w_{m+2}-w_{m+1} & w_{m+2}-w_{m+3} &  &  &  & \\
\\
 & 2(w_0-w_{2m+3}) & 0 & w_{m+2}-w_{3m+5} & & 0 &\\
 \\
 &  & 2(w_0-w_{2m+5}) & 0 & w_{m+2}-w_{3m+7} & & \\
 \\
 & *  &  & 2(w_0-w_{4m+7}) & 0 & w_{m+2}-w_{5m+9} & \\
 \\
 &  &  & & \ddots & \ddots & \ddots \\
\end{pmatrix}.
\end{eqnarray}
\end{widetext}
Fortunately, numerics indicates that the result converges really fast. Base on this observation, we find that the total valley Hall conductivity become
\begin{equation}
\begin{split}
8\sum_{i=b,c,d,e}\sigma_{xy}^{i,A}&=\frac{2me^2}{h}\cos\theta_{\bm k}\\
&\times[f_1(w_0-w_1)+f_{2m+3}(w_{m+2}-w_{m+1})].
\end{split}
\end{equation}
To gain some insight, we consider two special cases that may be relevant to experiments.

\subsection{$\epsilon=0$}

First we consider $\epsilon=0$, i.e., out-of-plane corrugations defined by Eq. (\ref{strain}) and (\ref{strain2}). By definition (\ref{omegam}), we have an iterative relation 
\begin{align}
w_{m}+w_{m-2}&=2w_{m-1}.
\end{align}
Since $w_{1}=w_0-1$, we obtain $w_{m}=w_0-m$. Then the matrix $T$ can be simplified, and the valley Hall conductivity reads
\begin{equation}
\begin{split}
8\sum_{i=b,c,d,e}\sigma_{xy}^{i,A}&=\frac{1.9662e^2}{h}\cos\theta_{\bm k}
\end{split}
\end{equation}
for $m=1$ (monolayer graphene) and
\begin{equation}
\begin{split}
8\sum_{i=b,c,d,e}\sigma_{xy}^{i,A}&=\frac{3.9634e^2}{h}\cos\theta_{\bm k}
\end{split}
\end{equation}
for $m=2$ (bilayer graphene). Note that these results are very close to $\frac{2me^2}{h}\cos\theta_{\bm k}$, differing by less than $2\%$, confirming the validity of the approximation used in Sec.~\ref{sec:diagram} and the main text.

\subsection{$\epsilon=0.821$}

In this section, we consider $\epsilon=0.821$, which is a more realistic value for thermally excited ripples~\cite{doussal1992}, i.e., out-of-plane corrugations. By definition (\ref{omegam}), we have an iterative relation
\begin{equation}
\begin{split}
w_{m}&=\epsilon\frac{1}{2(m-1)}(w_{m-2}-w_{m})\\
&+w_{m-1}-\frac{1}{2}w_{m-2}+\frac{1}{2}w_m,\   \ m\geq2
\end{split}
\end{equation}
and
\begin{align}
w_1=(\frac{1}{\epsilon}-1)w_0.
\end{align}
This leads to a solution
\begin{align}
w_{m}-w_0&=P_m(w_{1}-w_{0}),
\end{align}
where 
\begin{equation}
\begin{split}
P_m&=\frac{\prod_{ii=1}^{m-1}(ii-\epsilon)}{\prod_{ii=1}^{m-1}(ii+\epsilon)}
+\frac{\prod_{ii=1}^{m-2}(ii-\epsilon)}{\prod_{ii=1}^{m-2}(ii+\epsilon)}\\
&+\cdots+\frac{\prod_{ii=1}^{1}(ii-\epsilon)}{\prod_{ii=1}^{1}(ii+\epsilon)}+1.
\end{split}
\end{equation}
By numerics, we find that 
\begin{equation}
\begin{split}
8\sum_{i=b,c,d,e}\sigma_{xy}^{i,A}&=\frac{1.9994e^2}{h}\cos\theta_{\bm k}
\end{split}
\end{equation}
for $m=1$ and 
\begin{equation}
\begin{split}
8\sum_{i=b,c,d,e}\sigma_{xy}^{i,A}&=\frac{3.9998e^2}{h}\cos\theta_{\bm k}
\end{split}
\end{equation}
for $m=2$. Again the results are very close to $\frac{2me^2}{h}\cos\theta_{\bm k}$, differing by less than $0.1\%$.

\end{document}